\documentclass[numberedappendix]{emulateapj}

\usepackage{multirow}
\usepackage{graphicx}
\usepackage{rotating}
\usepackage{color}

\shorttitle{Timing analysis of SGR J1550-5418}
\shortauthors{Huppenkothen et al.}

\begin{document}

\title{Quasi-Periodic Oscillations in Short Recurring Bursts of  the Soft-Gamma Repeater J1550-5418}

\author{D. Huppenkothen\altaffilmark{1, 2}, C. D'Angelo\altaffilmark{1}, A. L. Watts\altaffilmark{1},  L. Heil\altaffilmark{1}, M. van der Klis\altaffilmark{1}, A. J. van der Horst\altaffilmark{1}, C. Kouveliotou\altaffilmark{3,4}, M.G. Baring\altaffilmark{5}, E. G{\"o}{\u g}{\"u}{\c s}\altaffilmark{6}, J. Granot\altaffilmark{7}, Y. Kaneko\altaffilmark{6}, L. Lin\altaffilmark{6}, A. von Kienlin \altaffilmark{8}, G. Younes\altaffilmark{4,10}}

\altaffiltext{1}{Astronomical Institute ``Anton Pannekoek'', University of
  Amsterdam, Postbus 94249, 1090 GE Amsterdam, the Netherlands}
\altaffiltext{2}{Email: D.Huppenkothen@uva.nl}
\altaffiltext{3}{Astrophysics Office, ZP 12, NASA-Marshall Space Flight Center, Huntsville, AL 35812, USA}
\altaffiltext{4}{NSSTC, 320 Sparkman Drive, Huntsville, AL 35805, USA}
  \altaffiltext{5}{Department of Physics and Astronomy, Rice University, MS-108, P.O. Box 1892, Houston, TX 77251, USA}

\altaffiltext{6}{Sabanc\i~University, Orhanl\i-Tuzla, \.Istanbul  34956, Turkey}
\altaffiltext{7}{Department of Natural Sciences, The Open University of Israel, 1 University Road, P.O. Box 808, Ra'anana 43537, Israel}
\altaffiltext{8}{Max-Planck-Institut f\"{u}r extraterrestrische Physik, Giessenbachstrasse 1, 85748 Garching, Germany}
\altaffiltext{9}{François Arago Centre, APC, 10 rue Alice Domon et Léonie Duquet, 75205 Paris, France}
  \altaffiltext{10}{Universities Space Research Association, 6767 Old Madison Pike, Suite 450, Huntsville, AL 35806, USA}

\begin{abstract}
This is an abstract. 
\end{abstract}
\begin{abstract}
The discovery of quasi-periodic oscillations (QPOs) in magnetar giant flares has opened up prospects for neutron star asteroseismology. The scarcity of giant flares makes a search for QPOs in the shorter, far more numerous bursts from Soft Gamma Repeaters (SGRs) desirable. In \citet{huppenkothen13}, we developed a Bayesian method for searching for QPOs in short magnetar bursts, taking into account the effects of the complicated burst structure, and have shown its feasibility on a small sample of bursts. Here, we apply the same method to a much larger sample from a burst storm of 286 bursts from SGR J1550-5418. We report a candidate signal at $260 \, \mathrm{Hz}$ in a search of the individual bursts, which is fairly broad. We also find two QPOs at $\sim 93 \, \mathrm{Hz}$, and one at $127 \, \mathrm{Hz}$, when averaging periodograms from a number of bursts in individual triggers, at frequencies close to QPOs previously observed in magnetar giant flares. Finally, for the first time, we explore the overall burst variability in the sample, and report a weak anti-correlation between the power-law index of the broadband model characterising aperiodic burst variability, and the burst duration: shorter bursts have steeper power law indices than longer bursts. This indicates that longer bursts vary over a broader range of time scales, and are not simply longer versions of the short bursts.   
\end{abstract} 

\keywords{pulsars: individual (SGR J1550-5418), stars: magnetic fields, stars: neutron, X-rays: bursts, methods: Time-series analysis, methods:statistics}

\section{Introduction}
Soft Gamma Repeaters (SGRs) represent a small class of neutron stars whose slow spin periods and high spin-down rates imply an unusually strong magnetic field in the excess of $10^{14} \, \mathrm{G}$. \citet{duncan1992} and \citet{1995MNRAS.275..255T} predicted the existence of such objects, which they named magnetars. SGRs are believed to be one of two observational manifestation of neutron stars with an exceptionally strong magnetic field; Anomalous X-ray Pulsars (AXPs) form the other class of objects, although evidence suggests that there is no clear-cut line between them, and recently a low-magnetic field source has been found \citep{2010Sci...330..944R}.

The defining characteristic of SGRs are irregular bursts that vary in duration from tens to hundreds of milliseconds and span $\sim 5$ orders of magnitude in peak luminosity ($10^{38}$ to $10^{43} \, \mathrm{erg} \, \mathrm{s}^{-1}$) in hard X-rays $< 100$ keV. 
However, there is a very rare type of burst, the so called giant flares, which have been only detected three times in the last $34$ years from three different sources. These reach peak luminosities of $\sim10^{45} \,\mathrm{erg}\,\mathrm{s}^{-1}$ and are believed to be powered by a catastrophic reordering of the magnetic field. Since this field is coupled to the solid crust, \citet{1998ApJ...498L..45D} suggested that such large-scale reconfiguration might rupture the crust, triggering global seismic vibrations that would be visible as periodic modulations of the X-ray and $\gamma$-ray flux. This idea was confirmed by the detection of quasi-periodic oscillations (QPOs, i.e. stochastic processes that vary on a characteristic time scale) in the expected range of frequencies  ($\sim 10-1000$ Hz) in the tails of giant flares from two different magnetars \citep{2005ApJ...628L..53I, 2005ApJ...632L.111S, 2006ApJ...653..593S, 2006ApJ...637L.117W}. SGR giant flares thus present outstanding text cases for testing theories of neutron star structure and composition models.
Several intermediate flares, in energy and duration between the short bursts and the giant flares, have also been observed, but no QPOs have been found in these bursts \citep{Watts11}.

To date, there have been few searches for QPOs in recurrent bursts of magnetars. \citet{2010ApJ...721L.121E} reported QPO detections in a sample of bursts from SGR 1806-20 observed between $2$ keV and $60$ keV with the {\it Rossi} X-ray Timing Explorer (RXTE), however, a revised analysis incorporating variability in the burst envelope showed that the reported QPOs are not significant \citep{huppenkothen13}.

Finding QPOs in short SGR bursts is technically challenging: as shown in \citet{huppenkothen13}, standard Fourier methods commonly used for this purpose fail when applied to the short, highly variable burst light curves. The major difficulty lies in the non-stationarity of a magnetar bursts. The statistical distributions generally used in Fourier analysis in astronomy are strictly only valid for processes whose properties do not vary over the duration of an observation. This is clearly not true for an SGR burst: they are short events, exhibiting variability on time scales roughly equivalent to the periods of QPOs observed in the giant flares. Below $100 \, \mathrm{Hz}$ or so, many of the bursts exhibit a wealth of variability properties: to leading order, there is the rise and fall of the burst, i.e. a burst envelope. In most bursts, the envelope has a high degree of temporal substructure beyond this envelope. This substructure differs widely from burst to burst, and is poorly understood. Perhaps what we call a burst is actually a superposition of many smaller events. Alternatively, the overall burst shape could be composed of an envelope combined with a stochastic process, leading to additional variability on shorter time scales. Finally, the complexity of the burst envelopes in general varies with energy, inserting another constraint in our interpretation of their structure. 
This lack of knowledge leads to two major problems when searching for QPOs. At low frequencies, very few cycles of a potential oscillation are sampled due to the short duration of the burst. A succession of peaks may look like a quasi-periodic signal to the naked eye, but could be a chance superposition of a stochastic process, without the characteristic timescale implied by a QPO.
The other major difficulty is our lack of knowledge of the underlying statistical distribution that we must test against. The statistical distributions generally used in testing for QPOs are strictly defined for stationary stochastic processes. The presence of a burst envelope changes the observed distributions at low frequencies from those we know. This makes it difficult to derive inferences about the presence of a QPO at these frequencies.

In the absence of this knowledge, it is possible to make reasonable assumptions. In \citet{huppenkothen13}, we introduced a Bayesian approach to deal with our uncertainty in the underlying burst processes by assuming a purely stochastic process with a power-law power spectrum, a so-called red noise process. While this assumption is strictly not true, either, we showed that it is a conservative choice: in practice, the presence of the burst envelope narrows the statistical distributions at low frequencies compared to the distribution we use to model the process. We are thus more likely to underestimate the significance of a signal at low frequency than overestimating our confidence in a detection.
In \citet{huppenkothen13}, we also analysed a short bursting episode of the magnetar SGR J0501+4516, where we found one candidate detection out of $27$ bursts. Our results were inconclusive with regard to the origin of this signal, and showed where our method can potentially produce ambiguous results: the significant detections were all at integer multiple frequencies of a low-frequency signal at $\sim 30 \, \mathrm{Hz}$, which was heavily affected by red noise and thus not significant itself. However, this signal corresponds to less than two full cycles at $30 \,\mathrm{Hz}$, given the short duration of the burst. We thus concluded that this signal was equally likely to be a chance occurrence of two red noise peaks close together as it was to be a QPO, and deferred a more in-depth discussion to a later work with a larger sample of bursts.

In this paper, we perform a comprehensive search for QPOs in a much larger sample of bursts from a so-called burst storm observed from SGR J1550-5418 in January 2009.
SGR J1550-5418 (also 1E 1547.0-5408) was first observed with the {\it Einstein} X-ray observatory \citep{lamb81}. Later observations with XMM-Newton revealed a soft X-ray spectrum and a possible association with a young supernova remnant suggesting that it might be an Anomalous X-ray Pulsar \citep[AXP][]{gelfand07}. The AXP nature was confirmed  by the subsequent detection of radio pulsations with a slow spin period of $P = 2.096\mathrm{s}$ and a spin-down of $\dot{P} = 2.318 \times 10^{-14}$, implying a magnetic field of $3.2 \times 10^{14} \, \mathrm{G}$ \citep{Camilo07}. 

SGR J1550-5418 exhibited three major bursting episodes: in October 2008, January 2009 and March/April 2009. The January episode was exceptional: the source showed hundreds of bursts within a single day, observed with several X-ray telescopes: the {\it Swift} Burst Alert Telescope (BAT) \citep{israel10, scholz11}, the {\it Fermi} Gamma-Ray Burst Monitor (GBM) \citep{kaneko10,vonkienlin12,vanderhorst12}, the {\it Rossi X-ray Timing Explorer} (RXTE) \citep{dib12} and two main instruments on board the {\it INTEGRAL} spacecraft \citep{mereghetti09, savchenko10}).

Burst storms like the one observed from SGR J1550-5418 are rare, and have been observed in only three other sources \citep[SGR 1806-20, SGR 1900+14 and SGR 1627-41;][]{goetz2006,israel2008,1999ApJ...519L.151M}, the first two of which have also exhibited a giant flare.
During the first triggered observation recorded with {\it Fermi}/GBM on 22 January 2009, the source also showed an increase in persistent flux level up to $\sim 100 \, \mathrm{keV}$ \citep{kaneko10} for around $150$ seconds of intense bursting. A subsequent search for pulsations in this plateau of hard emission revealed a signal at the period of the neutron star, but no higher-frequency QPOs.
The bursting episode ended in April 2009, and there have been no subsequent bursts recorded since. A catalogue of magnetar bursts observed with {\it Fermi}/GBM is currently in preparation (Collazzi et al., in preparation).

Here, we present the first large scale robust QPO search from the 2009 January burst storm, observed with {\it Fermi}/GBM. In Section \ref{sec:data} we describe the sample in some detail, in Section \ref{sec:analysis} we give a very brief overview of the Bayesian technique used in the QPO searches. Finally, in Section \ref{sec:results} we present our results and interpret both the QPO searches as well as a characterisation of broadband variability in the bursts in Section \ref{sec:discussion}.

\section{Data}
\label{sec:data}

X-ray bursts from SGR J1550-5418 triggered {\it Fermi}/GBM a total of 55 times between 2009 Jan 22  and 2009 Jan 29, with 41 triggers on Jan 22 alone. Each trigger records data from $30\,\mathrm{s}$ before the trigger up until $300\,\mathrm{s}$ after the trigger. As a result, multiple (untriggered) bursts were observed in most triggers. \citet{vanderhorst12} identified a total of 286 bursts in this sample, which have time-tagged event (TTE) data available. The TTE data type has a time resolution of $2\mu\mathrm{s}$, needed for high-precision timing studies. We use data from the $12$ NaI detectors, whose energy range of $8$ keV to $4$ MeV is sufficient, since SGR bursts rarely exhibit radiation above $200$ keV. Additionally, we only used detectors with viewing angles to the source $< 60^{\circ}$, and checked whether the source was occulted by the spacecraft and the other instrument, the Large Area Detector (LAT). 

The sample is the same as in \citet{vanderhorst12}, and we use the burst durations, start times, and fluences from that paper in our analysis. We extracted TTE data between $8 \, \mathrm{keV}$ and $200 \, \mathrm{keV}$ around each burst, starting at $t_{\mathrm{start}} - 0.1 \times\mathrm{T}90$ (the burst duration, $\mathrm{T}90$, is defined as the time in which the central $90\%$ of the photons, starting at $5\%$ and ending at $95\%$, reach the detector) and ending at $t_{\mathrm{start}} + 1.1\times\mathrm{T}90$ in order to ensure the entire burst is within our data set.
The sample has a mean duration of $0.174\,\mathrm{s}$ and an overall asymmetric shape with a faster rise than decay. The estimated fluences range from $10^{-8}$ to $10^{-5} \, \mathrm{erg}\, \mathrm{cm}^{-2}$. For a more detailed description of the data extraction process and sample definition, see \citet{vanderhorst12}. 
An analysis of the first trigger, including a timing analysis of the inter-burst periods, was performed in \citet{kaneko10}. Time-resolved spectroscopy and the spectral evolution with burst flux is discussed in \citet{younes13}.
Of the bursts in the sample, 23 have saturated parts, where the detector cannot record all photons during periods of very high count rates. We excluded all 23 bursts from our analysis due to the rather complicated effect saturation has on the timing analysis. This gives us a total sample of 263 bursts.

\section{Analysis Methods}
\label{sec:analysis}

QPOs are generally found by taking the Fourier transform of a light curve and looking for variability focussed at a particular frequency. The square of the Fourier transform of the data is called the {\it periodogram}\footnote{Note that throughout this paper, we use the term {\it periodogram} to denote the observed squared Fourier spectrum of a light curve, and use the term {\it power spectrum} for the underlying (potentially stochastic) process that may have produced the observed data}. Different types of variability have different frequency distributions. Our task becomes to disentangle the different components in the periodogram. While pure photon counting noise has a flat power spectrum with a well-behaved $\chi^2$ distribution with two degrees of freedom about a constant mean, QPOs produce sharp coherent features. Stochastic processes with correlated frequencies, often termed ``red noise'' or ``$1/f$ noise'', are also often observed, and follow power laws or broken power laws with stronger variability at low frequencies, and decreasing power at higher frequencies. 

The short duration of magnetar bursts means that this `low-frequency' variability has timescales similar to those of the QPOs observed in the giant flares. Thus, we must test for QPOs in a periodogram consisting of complicated variability. 
We adopt the method from \citet{huppenkothen13}, first suggested for red-noise dominated periodograms in \citet{vaughan2010}. This method assumes an exponential distribution of powers about the underlying power spectrum, which we assume to be a power law or broken power law.
To find QPOs, we fit a broadband noise power spectrum to each burst, which is then divided out. The highest outlier in the residuals is our candidate QPO detection. We then simulate fake periodograms using the broadband noise power spectrum and incorporating uncertainties in the model parameters, and perform the same detection procedure on those simulations. We can thus compare the observed highest data/model outlier with a distribution of data/model outliers from the simulations, to infer the probability that the observed outlier is a significant QPO.

Below, we give a very brief overview of the QPO search strategy, and refer the reader to \citet{huppenkothen13} for a detailed description, discussion of the method's limitations and tests on both simulated data and a smaller sample of magnetar bursts.

In more detail, for each burst: 

(1) We fit both a power law and a broken power law, i.e. the broadband noise model, to the periodogram of the burst observation. We fit the unnormalised posterior predictive distribution, consisting of a likelihood function following a $\chi^2$ distribution around the broadband noise model and priors that are independent of each other and of the form $p(\theta) = 1/\theta$ (scale prior) for scale parameters (e.g. broadband noise amplitudes) and flat otherwise. As a result, we obtain the so-called maximum-{\it a-posteriori} (MAP) as a result of the numerical optimisation step. The MAP estimate is the Bayesian equivalent of the maximum likelihood. For both models, we then construct the ratio of likelihoods at the parameter values corresponding to the MAP estimate.

(2) We sample the posterior predictive distribution of the simpler broadband noise model - the power law - using a Markov Chain Monte Carlo (MCMC) technique, in this case employing an affine invariant MCMC ensemble sampler \citep{goodman10}, as implemented in python by {\it emcee} \citep{2012arXiv1202.3665F}. The resulting ensemble of parameter values will follow the posterior distribution of the assumed broadband model, thus allowing for statistical inferences over this distribution. 

(3) We simulate $N_s$ artificial periodograms from the MCMC sample, and fit each with the two broadband noise models considered such that we can construct the likelihood ratio for each of the fake periodograms. This will allow us to construct a distribution of likelihood ratios from a sample we know to be derived from the simpler model. If the likelihood ratio obtained for the observed periodogram is an outlier of the distribution of likelihood ratios, then the observed data is unlikely to be generated from this model. Note that this is strictly evidence {\it against} the power law model; it is not direct evidence {\it in favour} of the more complex model. We use this approach to reject the simple power law model for cases where the posterior predictive $p$-value (the ratio of samples in the posterior distribution of likelihood ratios lying above the observed values, divided by the total number of samples in this distribution) falls below $0.05$. If this is true, we use the broken power law to model the broadband component of the periodogram; otherwise the simple power law model is adopted.
A threshold of $p < 0.05$ is not very stringent, but desirable. We would rather reject the simpler broadband noise model in favour of a more complex one. It is preferable to overfit the periodogram, rather than underfit, because broadband noise features not adequately modelled by the broadband noise model may instead be mistaken for QPOs in the subsequent analysis.

(4) We construct a second MCMC sample from the adopted broadband noise model in the same fashion as in step (2). We simulate $N_s$ periodograms from this sample and fit with the adopted broadband model. For the observed periodogram and each fake periodogram, we divide the periodogram by the best-fit MAP parameter estimate and define the test statistic $T_R = \mathrm{max}_j(2 I_j/S_j)$, where $I_j$ is the observed power $I$ at frequency $j$ and $S_j$ is the value of model $S$ at frequency $j$. This test statistic is the maximum power in the residual periodogram after dividing out the broadband noise model. In the ideal case where the parameters $\theta$ defining the model $S$ are perfectly known, $2 I_j/S_j$ follows a $\chi^2_2$ distribution. In reality, there is an uncertainty in $S_j$, since the parameters $\theta$ are not known exactly, leading to a deviation in the distribution of $T_R$ from the theoretical expectation. Sampling the posterior probability distribution of the parameters given the data and the model via MCMC allows us to construct the actual distribution of $T_R$ from simulated periodograms, taking into account all relevant uncertainties. We can thus construct a posterior distribution for the $T_R$ statistic under the null hypothesis that the observed maximum power is due to a stochastic aperiodic process. Comparing $T_R$ from the real data to this simulated distribution allows us to define a posterior predictive $p$-value for this null hypothesis. If the latter is small, then the observed maximum power is unlikely to be a product of a $\chi^2_2$ distribution. 
 
Although the giant flare QPOs were all very narrow, we cannot exclude broader signals in the shorter bursts. We search for these signals by performing exactly the same analysis as in step (4) on periodograms that are binned in frequency, where the bin widths are chosen from a  logarithmic scale between $1\, \mathrm{Hz}$ and $200 \, \mathrm{Hz}$. Binning a broad QPO signal makes it easier to detect, since the QPO is grouped into a single bin, while random fluctuations from one frequency bin to the next are suppressed. We search for QPOs in the same way as in the unbinned periodogram: by defining $T_R$ for the binned periodogram and comparing to the distribution of binned $T_R$ values from the sample of simulated periodograms.
In this case the $p$-values for different bins are not independent. To avoid excessive false positives, we accept significant detections only if they are detected in at least two different bin widths at the same frequency. 
In order to conserve computation time, we set the number of simulations $N_s = 10^{4}$. This implies that the significance can only be quoted to $p = 10^{-4}$ for a single trial. The detection limit we use depends on the number of trials: the more periodograms we search, the more likely it becomes to make a significant detection purely by chance, even if no signal is present. We thus require a more stringent detection limit for searching individual bursts, where we search hundreds of periodograms, than for searching averaged periodograms, where we only search $10$. For searching individual bursts as in Section \ref{sec:individualbursts}, we require $p < 10^{-4}$ for a single trial, corresponding to $p < 0.0263$ or roughly $2.3\sigma$, given the number of bursts in our sample. For the $10$ averaged periodograms we search in Sections \ref{sec:avgduration} and \ref{sec:avgobservation}, we choose $6 \times 10^{-3}$, or roughly $3.5\sigma$. All $p$-values given below are trial-corrected: the in the search of individual bursts by the total number of those bursts, i.e. $263$, and for the averaged periodograms by $10$, the number of averaged periodograms searched. The number of frequencies and bin widths we searched over is automatically taken into account by our methodology, by searching over the entire frequency and bin width range for the simulations as for the real data.

\section{Results}
\label{sec:results}

We searched light curves from 263 individual bursts for periodic signals and QPOs. In order to be sure to include the entire burst, we added $10\%$ on either side of the burst duration ($\mathrm{T}90$). Additionally, we constructed averaged periodograms from samples of bursts explore whether a signal could be re-excited in several bursts. Finally, we characterised broadband variability for the sample as a whole, which may guide future work on emission and trigger mechanisms.

\begin{figure*}[htbp]
\begin{center}
\includegraphics[width=18cm]{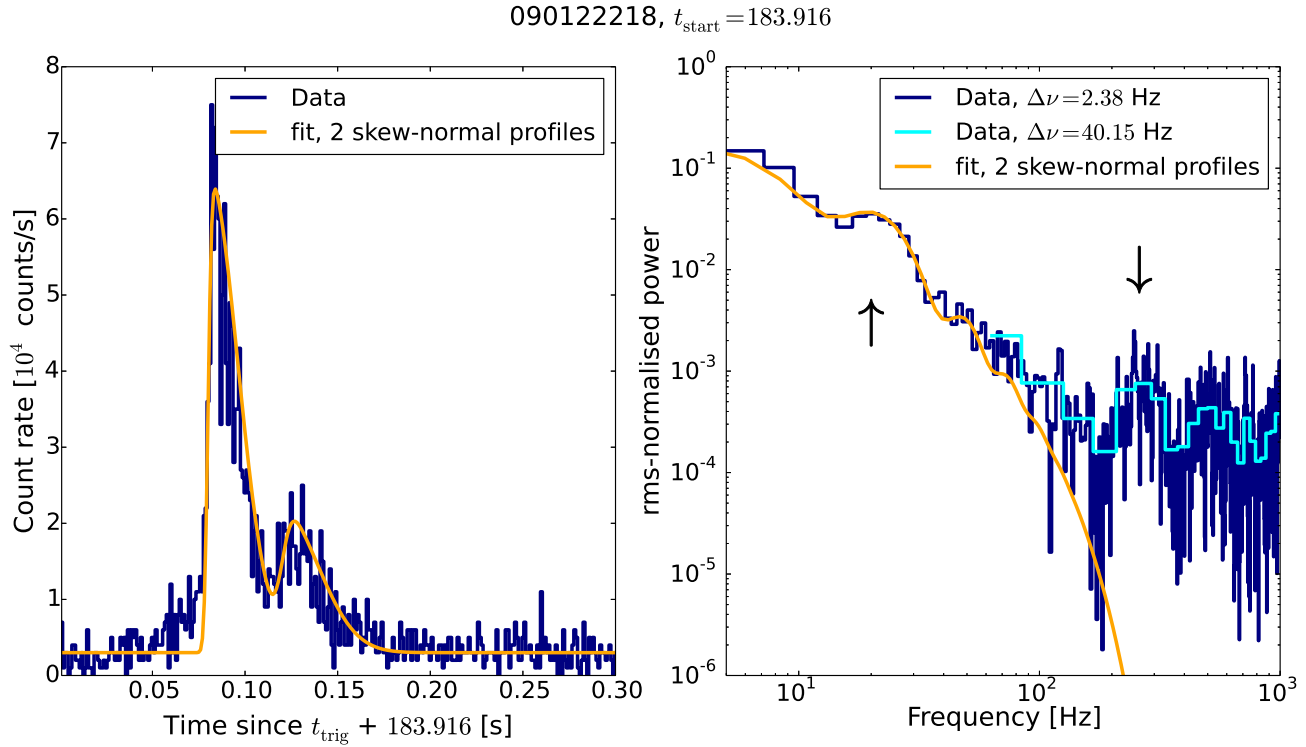}
\caption{Light curve (left) and periodogram (right) at two different frequency resolutions for a burst in TTE data of trigger $090122218$. There is a feature at $\sim 20 \, \mathrm{Hz}$, which can be explained by the superposition of two individual peaks, modelled with the skew-normal function of Equation \ref{eqn:skewnormal}. A second feature at $\sim 260 \, \mathrm{Hz}$ is significant ($p < 0.0263$) in the binned periodogram (in cyan on the same plot), but very broad, with a $Q$-value $Q = \nu/\Delta\nu = 2.9$. We added arrows to guide the eye.}
\label{fig:plotsig1}
\end{center}
\end{figure*}

\subsection{Individual Burst Searches}
\label{sec:individualbursts}

We searched all 263 bursts for QPOs over the complete range of available frequencies from $\leq 10 \, \mathrm{Hz}$ to $4000 \, \mathrm{Hz}$.
The maximum frequency was chosen to maximise computational efficiencies, while at the same time, oscillatory modes are unlikely to occur at a much higher frequency.

Four bursts show detections significant with $p < 10^{-4}$ (single-trial) or $p < 0.0263$ corrected for $N_b = 263$ trials. 
Three candidates are significantly affected by dead time and pile-up, that is, their count rate is close to the saturation count rate. This is the case when a significant number of photons arrive within less than $2.6\mu\mathrm{s}$ of each other (the dead time of the GBM recoding system), and are consequently recorded as a single photon. Here we used the highest intrinsic time resolution data from GBM: Time Tagged Event (TTE) data with $2\mu\mathrm{s}$ resolution. While a $2.6\mu\mathrm{s}$ dead time scale corresponds to a higher frequency than we are interested in, the above effects can nevertheless influence the periodogram in nontrivial detector-dependent ways, which are not retrievable or quantifiable. A proper treatment of affected bursts is beyond the scope of this work; we thus consider the QPO search on these bursts as inconclusive, and make no further statements about their properties.

The remaining burst, one of several in TTE data of trigger $090122218$ with a burst duration of $0.49 \, \mathrm{s}$, has a significant detection of a broad feature at $260 \, \mathrm{Hz}$ with $p < 0.0263$ (trial-corrected; also includes an uncertainty in the parameters of the broadband model). We plot the light curve and periodogram of this candidate in Figure \ref{fig:plotsig1}. While there might be some red noise power left at these frequencies, the signal is largely dominated by white noise. We use the traditional (analytical) test against white noise for an upper limit on the detection probability \citep{groth75,1989tns..conf...27V}. This would be a precise estimate if there was no red noise in the signal, but as we cannot exclude some contamination from red noise, this must be regarded as an upper limit instead. We find that the probability that the observed peak in the periodogram is due to Poisson counting noise alone is $p = 5.26 \times 10^{-6}$. The fractional rms amplitude is high, $\mathrm{rms}_\mathrm{frac} = 21\% \pm 3\%$, as estimated from integrating over the noise-level subtracted periodogram. We estimated the error following \citet{heil2012}. This error calculation is somewhat too simplistic for the periodogram we consider here: there may be a residual contribution of aperiodic variability contaminating the powers we integrate over, which is not taken into account properly. However, our lack of knowledge about the burst processes involved preclude us to run simulations to establish the error to a higher degree of precision.

We measured the $Q$-value, defined as the centroid frequency divided by the width of the signal, by comparing the $p$-values for periodograms of this burst binned at several frequency resolution, and picking the frequency resolution that yielded the lowest $p$-value to reflect the most likely width of the signal. The QPO is extremely broad: the $Q$-value is $Q = \nu_0/\Delta\nu = 2.9$. This is at the lower boundary of what one would call a QPO ($Q > 2$) as opposed to a broadband noise feature.  Due to the single occurrence of this signal in the sample, it is not possible to average periodograms, as is usually done to improve signal to noise and estimate errors, such that a Lorentzian fit to the feature is possible for a precise estimate of the width and the rms amplitude \citep{2006csxs.book...39V}. 

The periodogram of the same burst also shows a broad feature at 20 Hz. In order to understand the origin of this feature and its connection to the QPO at $260 \, \mathrm{Hz}$, we fit two fast-rise, slow-decay profiles to the light curve. We used the skew-normal distribution, a generalisation of a simple Gaussian profile that allows for skewness and has the form:

\begin{equation}
\label{eqn:skewnormal}
f(t) = \frac{2}{\sigma} \phi\left( \frac{t-\mu}{\sigma} \right) \Phi\left( \alpha \left( \frac{t - \mu}{\sigma} \right) \right) \; ,
\end{equation}

where 

\[
\phi\left( \frac{t - \mu}{\sigma} \right) = \frac{1}{\sigma \sqrt{2\pi}} \exp{-\frac{(t-\mu)^{2}}{2\sigma^{2}}} 
\]

and

\begin{equation}
\Phi\left( \alpha \left( \frac{t - \mu}{\sigma} \right) \right)  =  0.5 \left[ 1 + \mathrm{erf}\left( \alpha\frac{t - \mu}{\sqrt{2}\sigma}\right)  \right] \; .
\end{equation}

Here, $\mu$ is the location in time of the peak of the profile, $\sigma$ is the width and $\alpha$ a skewness parameter \citep{azzalini85}.
We find that the signal at $20$ Hz is easily reproduced by a superposition of two skewed peaks with a separation of $0.04\, \mathrm{s}$ and widths $\sigma_1 = 0.02\, \mathrm{s}$ and $\sigma_2 = 0.016\, \mathrm{s}$. While the feature is easily reproduced by two non-periodic functions, there are too few cycles observed to make a strong statement about its nature \citep[see][for a similar feature]{huppenkothen13}. However, it cannot explain the highly significant signal at $260\, \mathrm{Hz}$: the power spectrum of the two skew-normal functions fitted to the data turns over at lower frequencies and becomes negligible above $200 \, \mathrm{Hz}$. Beyond this frequency, there is very little power in this model, and the power spectrum at higher frequencies should be dominated by Poisson noise only. This implies that the QPO is not easily reproduced by a burst envelope, and is likely a separate process producing variability at these frequencies.
In order to confirm this observation, we have fit the observed light curve with both standard Gaussian profiles, as well as Lorentzian profiles. Both alternatives give results very similar to the one presented above: a near-perfect fit to the low-frequency feature and a sharp drop in power around $200\, \mathrm{Hz}$.

\begin{figure}[]
\begin{center}
\includegraphics[width=9cm]{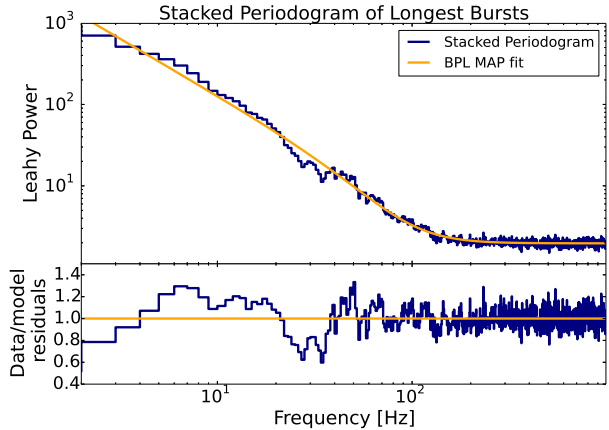}
\caption{Top: averaged periodogram (blue) of the 47 longest bursts, maximum a posteriori (MAP) fit of a broken power law to the periodogram (orange). The Leahy power is defined as $2|a_j|^{2}/N_{\mathrm{ph}}$, where $a_j$ is the Fourier amplitude at frequency $\nu_j$ and $N_{\mathrm{ph}}$ is the number of photons in a time series. Bottom: data/model residuals. The significant ($p < 2.5\times 10^{-3}$) signal is at $10 \, \mathrm{Hz}$, with a width of $\sim 5\, \mathrm{Hz}$. }
\label{fig:avgsig}
\end{center}
\end{figure}

The sensitivity limits for signal detection vary strongly from burst to burst and with frequency, especially for the low-frequency part of the periodogram where the contamination by broadband variability is strong. Below $\sim 100 \, \mathrm{Hz}$, sensitivies range from $\sim 50\%$ fractional rms amplitude at $30 \, \mathrm{Hz}$ to $\sim 10\%$ fractional rms amplitude at $100 \, \mathrm{Hz}$. 
Above $\sim 150 \, \mathrm{Hz}$, the bursts are almost all dominated by photon detector noise, and a QPO should be the only source of non-white-noise variability in this regime. Our method converges towards standard Fourier methods in this frequency range. Instrumental effects such as dead time can still be an issue; neither method is equipped to deal with these effects without a large number of dedicated simulations.  
Above $150 \, \mathrm{Hz}$, sensitivities are generally in the range of $5 - 10 \%$ fractional rms amplitude.

\subsection{Averaged Periodograms}
\label{sec:avgduration}
To increase sensitivity, we average the periodograms of a number of bursts. This assumes that the short bursts always excite the same star quakes, which has also been seen in giant flares, where QPOs are detected to be present over many cycles.

\subsubsection{Signal grouped by burst duration}

We sorted the bursts by duration (T90) into five groups: $<50\, \mathrm{ms}$, $50 - 100 \, \mathrm{ms}$, $100 - 250  \, \mathrm{ms}$, $250 - 500  \, \mathrm{ms}$, and $ > 500  \, \mathrm{ms}$. To average periodograms, we picked the longest burst in each group and extracted light curves of the same duration for each burst in the sample, so that each periodogram would have the same number of frequencies. We then averaged the periodograms within a group to get the final periodogram. Since we use light curves of equal duration within each group, the shorter bursts in each group add noise into the final averaged periodogram, which reduces the QPO detection threshold somewhat. Limiting this effect is our main reason for dividing the bursts into groups, so that we can search for QPOs in the longest bursts without a strong noise component added by including the shortest bursts in the same sample.
\begin{figure}[h!]
\begin{center}
\includegraphics[width=9cm]{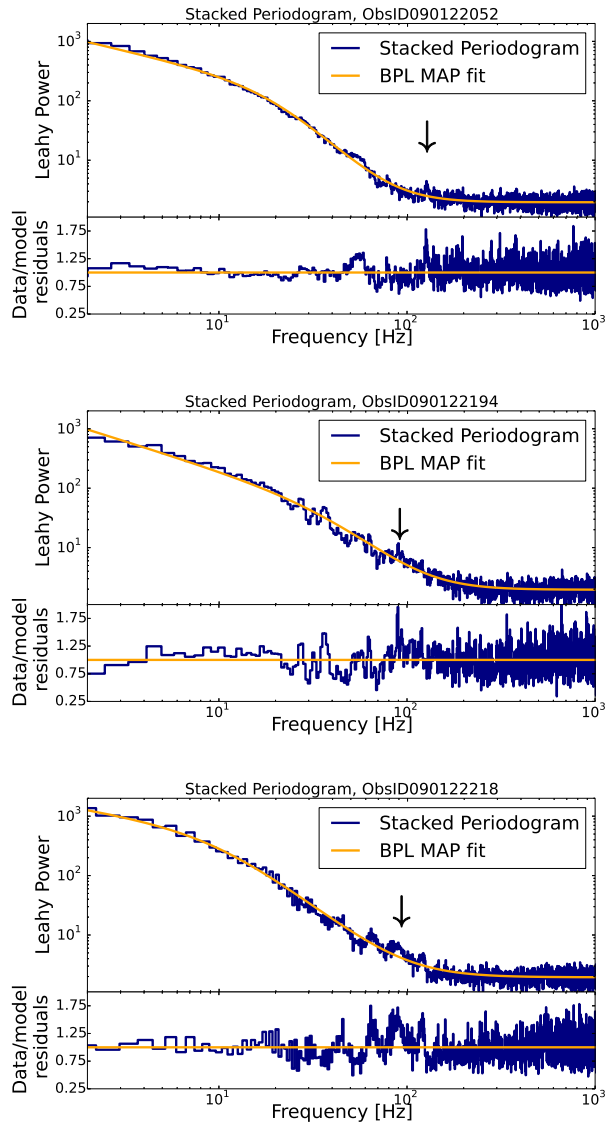}
\caption{Periodograms (blue, upper panels), MAP fits of a broken power law (orange) and data/model residuals (blue, lower panels) for the three triggers with candidate detections. Significant signals listed in Table \ref{tab:avgobs} are indicated with black arrows.}
\label{fig:avgobs}
\end{center}
\end{figure}

There are no QPOs detected in the first four averaged periodograms. We report a candidate detection in the averaged periodogram of the longest bursts (T90 $> 0.5\mathrm{s}$, $N_\mathrm{bursts} = 47$, see Figure \ref{fig:avgsig} for the averaged periodogram). The strongest signal with $p < 2.5 \times 10^{-3}$ occurs at $10 \, \mathrm{Hz}$, with a width of $\sim 5 \, \mathrm{Hz}$. Note that $10\, \mathrm{Hz}$ corresponds to a timescale of $0.1\, \mathrm{s}$, close to the peak of the distribution of burst durations. However, we cannot exclude that this feature is actually an artefact caused by an inadequate characterisation of the underlying power spectrum. Another process, such as a doubly-broken power law or a combination of Lorentzians as often used in broadband noise modelling of X-ray binaries, may represent the shape of the power spectrum better, but requires more intricate model selection criteria than implemented here.

\newpage
\subsubsection{Averaged Periodograms per Trigger}
\label{sec:avgobservation}

\begin{deluxetable*}{lcccccc|ccccc}
\label{tab:avgrms}
\tablewidth{500pt}
\tablecolumns{6}
\tablecaption{Signals from the five averaged triggers, and detection sensitivities for different frequencies}
\tablehead{
\colhead{Trigger ID} &
\colhead{$N_{\mathrm{bursts}}$} &
\colhead{min T90 [s]} &
\colhead{max T90 [s]} &
\colhead{$\nu_0$ [Hz]} &
\colhead{$\Delta\nu$ [Hz]} &
\colhead{posterior} &
\colhead{simulated}&
\multicolumn{4}{c}{Sensitivities in fractional rms amplitude}  \\
\colhead{} &
\colhead{} &
\colhead{} &
\colhead{} &
\colhead{} &
\colhead{} &
\colhead{$p$-value} &
\colhead{$p$-value}&
\colhead{40 Hz} &
\colhead{70 Hz} &
\colhead{100 Hz} &
\colhead{1000 Hz}}
 \startdata
 090122037	& 	$32$	& $0.0322$	& $1.0724$	&	$99$		&	$27$		& 	$< 4 \times 10^{-4}$ 		&	$0.107$		&	$3.6$	& 	$2.4$	& 	\nodata	&	$1.4$		 \\
 090122052	&	$28$	& $0.0364$	& $1.4952$	&	$127$	&	$10$		&	$< 4 \times 10^{-4}$		&	$0.016$		&	$4.8$	&	$2.5$	&	$1.9$	&	$1.5$		\\
 090122194	&	$20$	& $0.0364$	& $1.2124$	&	$93$		&	$12$		&	$< 4 \times 10^{-4}$ 		& 	$0.013$&	$6.5$	&	$3.8$	&	$2.7$	&	$1.8$			\\
 090122218	&	$21$	& $0.1176$	& $1.3496$	&	$91$		&	$10$		&	$1.2 \times 10^{-3}$ 	&	$0.009$		&	$5.2$	&	$3.0$	&	$2.3$	&	$1.6$	 \\
 090122283	&	$30$	& $0.0504$	& $2.4724$	&	$61$		&	$20$		&	$8 \times 10^{-3}$	&	\nodata &	$2.9$	&	$1.7$	&	$1.2$	&	$0.9$	
 \enddata
 \tablecomments{This table summarises the results from the averaged periodograms of five triggers. The significant detections are shown in Figure \ref{fig:avgobs}. The last, $090122283$ had no significant detections with $p < 6 \times 10^{-3}$, single trial probability. The second column gives the number of bursts averaged together, which equals the number of bursts in the trigger, excluding those that have saturated parts. The third and fourth column give the minimum and maximum burst durations in the sample. Columns five and six present the centroid frequency $\nu_0$ of the observed signal and the corresponding frequency bin width $\Delta\nu$ in which the signal is detected. We quote the detection threshold sensitivities where no detection has been made. The posterior $p$-value is estimated from simulations derived from the MCMC sample of the broadband model for each periodogram. The second p-value is derived from averaging random subsets of burst periodograms and extracting the highest outlier from the data-model residuals for each averaged periodogram, as described in the text.}
\label{tab:avgobs}
\end{deluxetable*}

We also search for QPOs in trigger data sets with high resolution TTE data. Since data sets obtained with {\it Fermi}/GBM from an individual trigger are roughly $330 \, \mathrm{s}$, we searched those triggers with a large number of bursts (see Table \ref{tab:avgobs} for an overview) in a short time span for long-lived signals. As for the duration-averaged periodograms, we extracted light curves around all bursts in those five triggers of the duration of the longest burst in that trigger data set. We then constructed the periodograms of these light curves and computed the average periodogram of the sample. The resulting periodograms do not all have the same frequency resolution; for those with a frequency resolution less than $1$ Hz, we averaged neighbouring frequency bins to achieve a resolution close to $1$ Hz.

We searched data sets from five triggers with $20$ to $32$ bursts per trigger, and excluded long timescale variability below $60 \, \mathrm{Hz}$ from the range of frequencies searched. Below $60 \, \mathrm{Hz}$, there will be a significant contribution from the overall shape of the short bursts (as in the $20\, \mathrm{Hz}$ feature discussed in Section \ref{sec:individualbursts}), and thus our estimates are unreliable. We search both the unbinned periodogram as well as periodograms binned to different frequency resolutions between $1\, \mathrm{Hz}$ and $200 \, \mathrm{Hz}$, but considered only candidate signals with $Q > 7$. This is necessary, because at low frequencies, a candidate signal in a frequency bin that is wider than $0.2\nu_0$ likely incorporates power from the part of the periodogram below $60 \, \mathrm{Hz}$ where we believe estimates to be unreliable.
We find candidate detections in three of the triggers. The results are summarised in Table \ref{tab:avgobs}. The signals, at $93$ Hz (trigger IDs $090122194$), at $91$ Hz (trigger ID $090122218$) and at $127$ Hz (trigger ID $090122052$), are fairly narrow, $Q \approx 7$ and $Q \approx 13$, respectively. Two of the signals are significant to $p < 4 \times 10^{-4}$ (roughly $3.7\sigma$), the third has a $p$-value of $p = 1.2 \times 10^{-3}$ ($3.4\sigma$). A fourth signal (trigger ID $090122037$) is significant to $p <4 \times 10^{-4}$, but fails to fulfil our criterion of $Q > 7$. At the same time, this signal is at a frequency of $99 \, \mathrm{Hz}$, close to the frequency where significant detections were made in two of the other triggers.   
We plot the periodograms for all three triggers in Figure \ref{fig:avgobs}. 
\begin{figure*}[]
\begin{center}
\includegraphics[width=18cm]{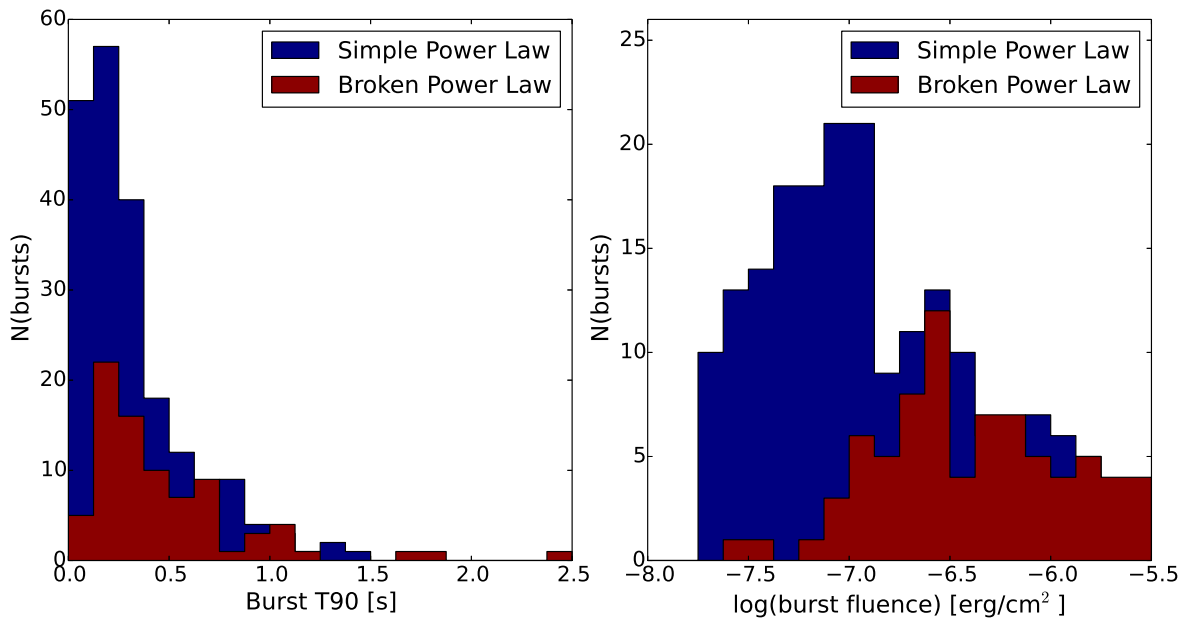}
\caption{Distributions of burst fluence and burst duration for the sample of bursts modelled by a simple power law (blue) and a broken power law (red). The burst duration is defined as the photon count T90. While there is only a marginal difference in T90 distributions between the two samples, there is a significant difference between the fluence distributions, $p = 9.32 \times 10^{-11}$: more complex bursts seem to have a higher fluence.}
\label{fig:fluenceduration}
\end{center}
\end{figure*}

The periodogram shape may change between different bursts, largely due to the wide spread in burst duration, fluence and burst shape. The effects this may have on the averaged periodogram are hard to quantify without a large number of dedicated simulations of the overall burst variability, which is beyond the scope of this work. In order to test whether the observed QPOs could be due to the differences in duration, fluence and shape in the averaged samples, we constructed a large number ($N_s=10^{3}$) of averaged periodograms from randomly selected subsets of the burst sample, excluding the four triggers where candidates were observed. If the QPOs are due to effects of the varying burst properties, then these signals should appear in a large number of these simulations. We searched these periodograms in the same way we did for the averaged periodograms from individual trigger data, and compared the resulting distribution of maximum powers $> 60 \mathrm{Hz}$ from the data-model residuals to the maximum powers from the averaged periodograms of individual triggers. Column 8 in Table \ref{tab:avgobs} shows the p-values of observing the candidate signals presented above from a random subset of bursts.

In Table \ref{tab:avgobs}, we also show the detection sensitivities for all five averaged periodograms for various frequencies, for the $\sim 1\, \mathrm{Hz}$ resolution of the periodogram. Note that we are even more sensitive to broader signals, as averaging neighbouring frequency bins increases the signal-to-noise ratio. The numbers quoted here are upper limits for the most coherent signal we could have observed, and were calculated for each quoted frequency from the distribution of maximum powers of the MCMC-derived sample of periodograms. 
 The frequency-dependence of the sensitivity is due to the fact the low frequency part of the periodogram is dominated by aperiodic `red noise' variability, and any quasi-periodic signal needs to introduce strong variations in order to be visible. Sensitivities for fractional rms amplitudes are between $3\%$ and $6\%$ for the lowest frequencies, and drop to $0.9\%$ to $1.7\%$ at high frequencies. The differences in sensitivities between triggers is due to a combination of number of averaged bursts, number of averaged frequencies, and the average count rates of bursts included.

\subsection{Broadband Variability}
\label{sec:broadbandnoise}

\begin{figure}[]
\begin{center}
\includegraphics[width=9.0cm]{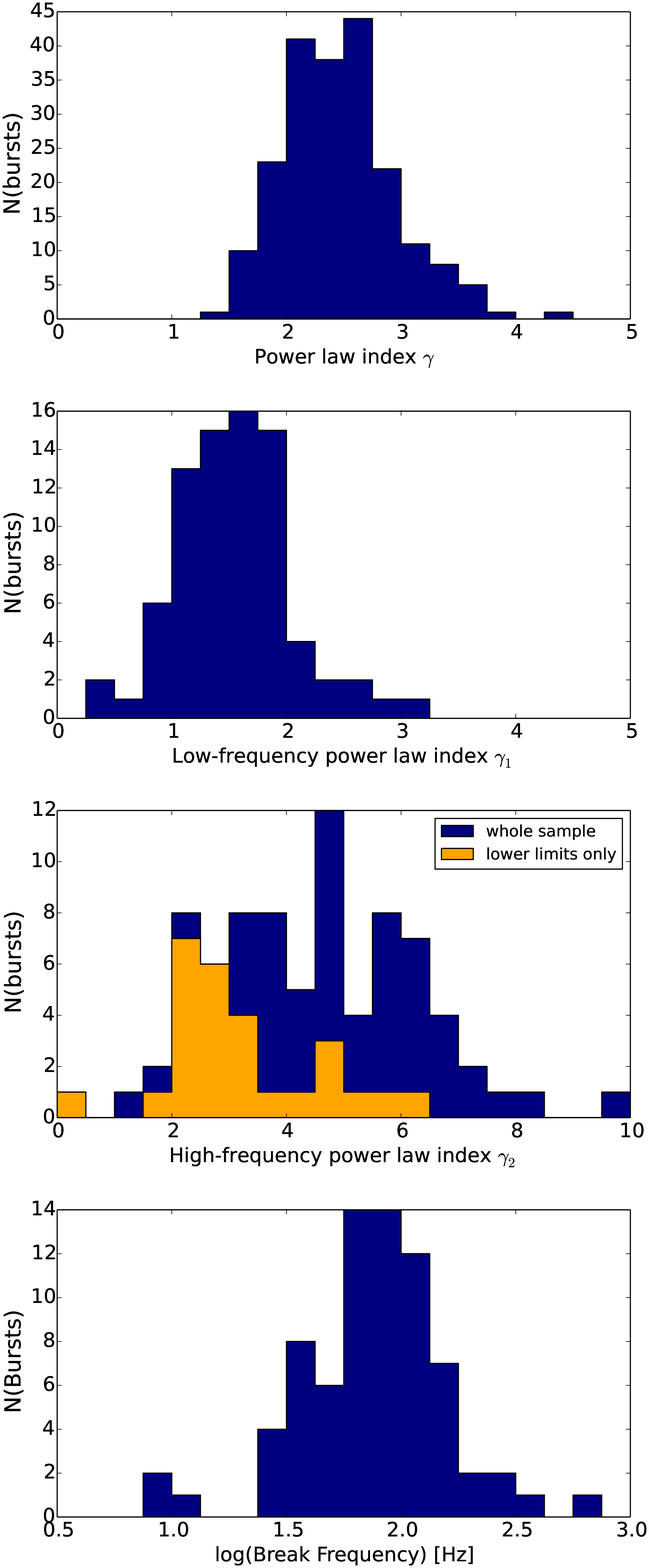}
\caption{Distributions for the power-law index for bursts modelled with a simple power law (top), the low-frequency power law index for bursts modelled with the broken power law (second from top), the high-frequency power law index for the latter model (third from top), and the break frequency between the low-frequency and high-frequency power law (bottom). Note that the high-frequency power law indices plotted here are the means of their posterior distributions, which, for some of the bursts, has a high variance. For the bursts where the  high-frequency power law index is largely unconstrained, we include the $5\%$ quantile in the sample instead, and include the fraction of these bursts in panel 3 in orange.}
\label{fig:gamma}
\end{center}
\end{figure}

\begin{figure}[]
\begin{center}
\includegraphics[width=9cm]{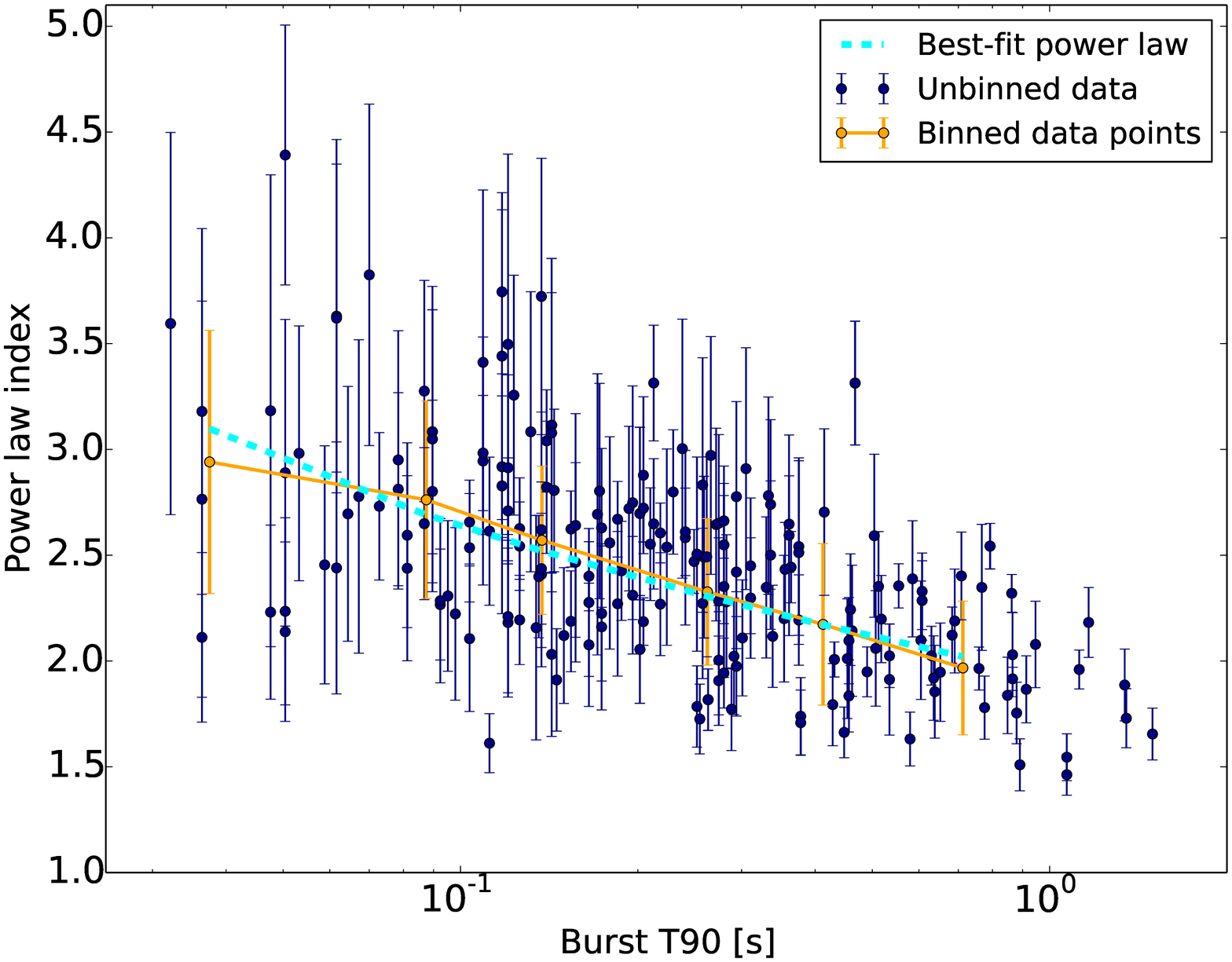}
\caption{Burst duration (T90) versus power law index for all bursts modelled with a simple power law. Errors on the power law index are estimated from the marginalised posterior distribution approximated with the MCMC sample. We show the data binned logarithmically and averaged within each bin in orange.  We tested for correlation using a Spearman rank coefficient, $R = -0.51$. The dashed, light blue line indicates a power-law fit ($\alpha = -0.1$) to the binned data points, as explained in the text.}
\label{fig:plindcorr}
\end{center}
\end{figure}

Magnetar bursts are a class of events with very diverse light curves: they differ vastly in duration and peak count rates, but also in overall burst shape \citep[see][for examples of burst light curves]{huppenkothen13}. How exactly this variability is produced is not well understood. Are all bursts realisation of fundamentally the same process? Are there characteristic rise or decay time scales? It is useful to characterise the variability properties of a large sample, which may answer some of these questions.

In the following, we give an overview of the broadband variability observed in the whole sample of bursts. Out of 263 burst periodograms, 193 were adequately fit with a simple power law plus a constant to account for the white noise component; the remaining 70 rejected the null hypothesis to $p < 0.05$, and we thus adopted a broken power law for these periodograms. Burst duration and burst fluence could influence whether a simple power law or a broken power law fits the broadband noise. For example, for dim bursts, variability observed in the bright bursts may be hidden in the noise. In order to test this idea, we plot the fluence and burst duration distributions for bursts modelled with a power law and a broken power law in Figure \ref{fig:fluenceduration}.  

Burst duration (T90) shows only a marginal difference in the T90 distribution ($p = 4.6 \times 10^{-4}$ for a two-sample Kolmogorov-Smirnov (KS) test). There is an appreciable difference in fluence between the samples (2-sample KS-test: $p = 9.32 \times 10^{-11}$): bursts with broken power law power spectra have higher fluence than bursts modelled with a simple power law. Note that the threshold for rejecting the power-law broadband model is not very high, $p < 0.05$. This is desirable for the main objective of our analysis, the search for QPOs: if the broadband noise is not adequately represented by the model, then broadband features may be attributed to QPOs instead, leading to false positive detections. Setting a threshold of $5\%$ is a compromise between reliability of QPO detection at the expense of potentially contaminating our sample of bursts fit with a broken power law with bursts that are consistent with a simple power law. In practice, this decreases the probability for rejecting the null hypothesis when performing the KS-test: the difference between the two distributions may be stronger than we report here. 

As well as studying the overall properties of bursts with different broadband noise models, we can also study the broadband noise properties of the sample, and see whether the noise properties change with fluence or burst duration. In Figure \ref{fig:gamma}, we show the distribution of power law indices for the various components. For the bursts modelled by a simple power law, the distribution of power law indices varies between 1.5 and 4, with a median at $\mu_{\gamma} = 2.42$.  The average low-frequency component of the two-component broken power law is flatter than for a single power law ($\mu_{\gamma_0} = 1.49$), while the higher frequency component is much steeper ($\mu_{\gamma_1} = 6.16$). Note that for several bursts, the second component is extremely steep. This may be caused by the contamination of this sample with bursts that were incorrectly classified as too complex for the simple power law. In this case, the second power law index is often not well constrained and tends to high values.
In Figure \ref{fig:gamma}, we also plot the break frequency between the two components of the broken power law for those burst periodograms for which the simple power law was rejected. The distribution peaks around $100\,\mathrm{Hz}$, below which the power law index is fairly flat. At higher frequencies, it steepens considerably, as shown in panel 3 of Figure \ref{fig:gamma}. The distribution is fairly broad, with the bulk of burst periodograms breaking between $30\,\mathrm{Hz}$ and $400 \, \mathrm{Hz}$.

We correlated the power law indices with various burst properties to see whether there is a systematic effect due to burst duration or brightness, similarly to the reasoning for why some bursts require a more complex model than a simple power law. There is an indication of a correlation of the burst duration with the power law index $\gamma$ for bursts modelled with a simple power law (see Figure \ref{fig:plindcorr}): shorter bursts seem to have slightly steeper power law indices. A Spearman rank coefficient yields $R = -0.51$, indicating an anti-correlation with a probability for the null hypothesis (no correlation) of $p = 8.65 \times 10^{-16}$. 
In order to compute the slope of the correlation as well as incorporate the errors on the measurements of the power law index, we binned the data set logarithmically into $7$ bins and computed the mean power law index within each bin \citep[following work in e.g. in X-ray binaries and Active Galactic Nuclei,][]{gleissner04}. We estimated the error on the mean as a standard error, $\sigma/\sqrt{n}$, where $\sigma$ is the standard deviation of the sample and $n$ is the number of data points in each bin. The correlation can be fit with a power law, with index $\alpha = -0.1027 \pm 0.00523$.

In order to test whether the results presented in this section could be artefacts of systematic effects that we failed to take into account. We sampled randomly from the posterior distributions of the parameters of the broken power law model from all bursts where the null hypothesis was rejected. We then sampled the distributions of $\mathrm{T}90$ and fluence from the observed ensemble of bursts, and created fake light curves with the power spectral properties of the broken power law model with the randomly sampled parameter sets, and combinations of burst $\mathrm{T}90$ and burst fluence taken from the real sample. We simulated light curves following \citet{1995A&A...300..707T}, and included Poisson statistics in the simulated light curves. For 1000 such fake bursts, we fit the periodograms and performed model selection between the power law and broken power law models.
$707$ periodograms simulated from the broken power law model were actually adequately fit with a simple power law instead, whereas only $293$ bursts rejected the null hypothesis. This indicates that the fit is strongly dependent on burst duration and fluence. Especially for short bursts, where there are few frequencies below $100 \, \mathrm{Hz}$, there may not be enough data points to require a more complex model. 
The power law and broken power law samples are well separated both in terms of burst $\mathrm{T}90$ and burst fluence. A two-sided KS-test reveals a separation in burst duration, $p = 1.0 \times 10^{-40}$, much stronger than observed in the data. The separation between the two samples in terms of the fluence distribution is comparable to that observed from the sample, $p = 2.2 \times 10^{-10}$. This indicates the possibility that separating bursts by their preferred broadband model may not be meaningful: perhaps all bursts follow a broad-band noise process that is closer to a broken power law than a simple power law, but for short bursts, there are not enough data points at low frequencies to confidently reject a simple power law model. 

We also tested for the presence of a correlation between burst $\mathrm{T}90$ and power law index for the sample of fake periodograms that accepted a power law model. While there still seems to be a correlation (Spearman rank coefficient p-value for no correlation: $p = 1.35 \times 10^{-10}$), this correlation is completely flat, with a power law slope of $\alpha = 1.77 \times 10^{-10}$, centred around a mean power law index of $\langle \gamma \rangle = 2.56$.

\section{Discussion}
\label{sec:discussion}

We have searched both individual bursts and averaged periodograms from samples of bursts for QPOs. Our analysis is the most precise to date for fast transients while taking into account the effects of aperiodic variability, but it is also conservative: at low frequencies, real quasi-periodic features may be missed because we assume the burst is a purely stochastic process, when it is not. 
Additionally, we do not model several effects which can significantly affect the outcome of a QPO search. Dead time can significantly affect especially the bright bursts, and thus render inferences invalid even at high frequencies. At low frequencies, we have demonstrated that some of the power concentrated in broad bumps, which can be classified as QPOs with broad widths by the detection algorithm, can easily be modelled with a simple estimate of a burst envelope consisting of two functions with a faster rise than decay. We must hence be very careful when interpreting signals at frequencies comparable to the lowest QPO frequencies seen in the giant flares ($18 \, \mathrm{Hz}$ and $36 \, \mathrm{Hz}$). 

\subsection{Individual Bursts}

We find no indication in individual bursts for QPOs at the frequencies and coherences seen in the giant flares.
We detected one significant signal at $260$ Hz, in the same burst where we found the broad feature at $20$ Hz that the algorithm flagged as significant. The latter is just as likely to be caused by a superposition of two burst envelope profiles as it is to be a QPO. With less than two full cycles, it is impossible to tell both models apart. This is reminiscent of the candidate detection reported in \citet{huppenkothen13}, where we noted that this candidate could be due to a chance occurrence of two such peaks close together.
With a $Q$-value of $2.9$, the signal at $260 \, \mathrm{Hz}$ is far broader than anything seen in the giant flare QPOs ($Q > 10$), but very strong, with a fractional rms amplitude of $21 \pm 3 \%$. The burst is longer than average, $\mathrm{T}90 = 0.48\, \mathrm{s}$, with a fluence at the lower end of the sample, $F = 3.94 \times 10^{-7} \, \mathrm{erg}\, \mathrm{cm}^{-2}$. The detected QPO at $260 \, \mathrm{Hz}$ is not present in any other burst in the entire sample, nor is it seen in the averaged periodogram of all bursts in this trigger, as described in Section \ref{sec:avgobservation}.

\subsection{Averaged Bursts}
We find a candidate detection in the averaged periodogram of the longest bursts with durations $\mathrm{T}90 > 0.5 \, \mathrm{s}$. These bursts are highly structured and generally have multiple peaks. The detected signal at $10$ Hz is quite broad and matches the position of the maximum in the distribution of burst durations in \citet{vanderhorst12}. This suggests a characteristic time scale for individual peaks within highly structured bursts, rather than a crustal mode. This in turn raises the question of whether these many-peaked bursts are causally connected single events, or instead individual bursts that happen to appear close to each other. While our results favour the latter explanation, it has been argued that these should be causally connected events. One argument is based on the distribution of waiting times between bursts: while the waiting times between bursts generally follow a log-normal distribution, an excess of short waiting times has been observed when regarding each peak as an individual burst, rather than grouping these peaks into causally connected events \citep{gogus99, gogus00}. 
However, at this point, we cannot exclude that this candidate detection arose from an inadequate broadband noise model fit to the data. A more complex periodogram shape, with additional power law components, could explain the observed excess power. To check this would require a somewhat more complex set-up of the model selection procedure, or should ideally be done with methods better suited to model broadband variability. We thus defer this task to a future work. 

The most interesting results stem from averaging periodograms in individual triggers. This allows us to probe timescales on which QPOs were observed in the giant flares ($10$s to $100$s of seconds). We find signals in two of the averaged periodograms of five triggers, at a frequency of $93 \, \mathrm{Hz}$, very close to the strongest QPO reported in the 2004 giant flare at $92\, \mathrm{Hz}$ \citep{2005ApJ...628L..53I}. In the periodogram of bursts from a third trigger, we find a significant detection at a slightly higher frequency, $\nu_0 = 127 \, \mathrm{Hz}$, where no giant flare QPO has been observed. 

When deriving inferences from an averaged periodogram, one makes the implicit assumption that all periodograms used to construct the average are realisations of the same underlying process. This need not be true for SGR bursts: we are averaging periodograms from bursts with vastly different durations and morphologies. While most bursts are described well by simple power laws or broken power laws in the Fourier domain, the parameters of these broadband noise models vary from burst to burst, as does the range of frequencies over which variability is observed (as seen e.g. by the correlation between burst duration and power law index reported in Section \ref{sec:broadbandnoise}). Additionally, the results reported in Section \ref{sec:individualbursts} show that at low frequencies, the burst envelope may dominate the power spectrum, which significantly alters both the shape and statistical distribution of the periodogram. In order to test this effect, we created averaged periodograms from randomly selected bursts of the sample, excluding the triggers where we detected a significant signal in the averaged periodograms. If we see many significant signals of the observed strength in these averaged periodograms from randomly selected bursts, then either the QPO we are interested in is re-excited in many of the bursts, or there are effects due to averaging vastly different bursts that our broadband noise model cannot take into account properly. 
The former case is unlikely: if a QPO at $93 \, \mathrm{Hz}$ were present in many bursts, we would have likely observed it when averaging by duration, as in Section \ref{sec:avgduration}. All three signals observed at $\sim 93 \, \mathrm{Hz}$ and $127 \, \mathrm{Hz}$ are fairly narrow, and at comparatively high frequency. At $\sim 90 \, \mathrm{Hz}$, the contribution by the burst envelope should be small (cf. Figure \ref{fig:plotsig1}), and not cluster around a single frequency, but rather follow a power law. While the chance to observe one such signal is still too high to claim a strong detection ($p \approx 0.01$ for all three narrow candidates), the fact that it is observed twice out of five trials, at a frequency close to that observed in the 2004 giant flare ($\nu = 92 \, \mathrm{Hz}$), strengthens the claim for a detection.

We note that even for those frequencies we do not detect any signal, we can quote stringent sensitivities that set quite tight upper limits on a signal that could have been there and go undetected by our algorithm. In the white noise regime above $\sim 150 \, \mathrm{Hz}$ or so, where our algorithm approaches classical Fourier methods, the variability is lowest, and thus we have the highest sensitivity to weak signals. We can confidently exclude high-frequency QPOs at $625$ Hz and $1840$ Hz that have been observed in the giant flares \citep{2006ApJ...653..593S, 2006ApJ...637L.117W}. The QPO at $625$ Hz was observed over a large fraction of the tail of the giant flare ($>150 \, \mathrm{s}$) in two different energy ranges, with a high fractional rms amplitude of $8.5\%$. Conversely, the QPO at $1840$ Hz is seen only in two cycles, but with high significant and a large fractional rms amplitude of $18\%$. Since our sensitivities for all five averaged periodograms are much lower at these frequencies ($< 1.7\%$), we can exclude a QPO of this type in the smaller flares from the burst storm to high degree of confidence.

\subsection{Aperiodic Variability}

While the aperiodic variability in short magnetar bursts is a hindrance when searching for QPOs, it is interesting in its own right. Each burst has a unique temporal structure, which can sometimes be quite complex. Nevertheless, most burst periodograms can be modelled fairly well with simple empirical models, which allows us to draw a number of general conclusions, and give indications where further work is required. The simplest question one can ask about is the separation between bursts modelled with simple power laws and those that require a more complex model, in our case a broken power law. Our results indicate that the differences between the two samples are largely due to systematic effects: for short bursts, fewer data points in the low-frequency part of the periodogram makes it more difficult to constrain the shape of the power spectrum, and these bursts are thus more likely to accept a simple power law as model for the underlying power spectrum. Similarly, a burst with a lower fluence will be more strongly affected by photon detector noise, rendering inferences about the shape of the power spectrum more difficult. This idea is strengthened by the fact that the fraction of bursts fit by a broken power law in both the observed sample and the sample of fake periodograms simulated from the broken power law, are very close: $27\%$ of observed bursts are fit by a broken power law, versus $28\%$ bursts in the simulated sample.
There is one striking discrepancy between the data and the simulations: in the simulated ensemble, the samples fit by the different models are very strongly separated in burst duration, whereas there is only mild evidence for this separation in the observed ensemble. One reason may lie in the lower number of bursts in the observed sample. Another reason may lie in the nature of our simulations: we simulated light curves from the posterior distributions of broken power law parameters inferred for the real bursts following the method of \citet{1995A&A...300..707T}. This method provides pure red noise light curves, which are only an approximation to the real data, as discussed in \citet{huppenkothen13}. It is possible that the separation in the simulated samples in burst duration are due to effects that are not adequately captured by this model. Without better knowledge of the underlying processes involved, however, we cannot construct a model more representative of the observed data, in order to increase  confidence in our inferences.

We also extracted the distributions of broadband model parameters from the means of the MCMC samples of each individual bursts. Note that the mean is a very simple estimate of the posterior probability distribution of the parameter. It can encode neither a skewness in the distribution, nor correlations between parameters. The only way to encode the full information of both marginalised distributions of parameters as well as potential correlations between parameters is to plot the full posterior probability distribution, which, for three or more parameters, is impossible. For the purpose of this study, we accept the simple estimate and its limitations, and refer a more nuanced analysis to future work. 
In general, the power law index for bursts modelled by a simple power law is confined between $1.5$ and $4$. While the distribution is fairly broad, it peaks around $\mu_{\gamma} = 2.42$. This is much higher than seen, for example, in gamma-ray bursts, where a similar analysis yields indices of $1.7$ to $2.0$ \citep{guidorzi12, beloborodov00}.

The search for correlations reveals two interesting observations. There is no correlation between power law index and fluence for the bursts that can be modelled with a simple power law. This is not surprising: the highest-fluence bursts preferentially reject the simpler model, and are consequently in the sample of bursts modelled with a broken power law. 
Secondly, we find an anti-correlation of power law indices with burst duration: shorter bursts have steeper power laws. The anti-correlation can be modelled with a simple power law with $\alpha = -0.1$. The observed correlation is unlikely due to systematic effects: simulated periodograms indicate a distribution of power law indices centred around $\langle \gamma \rangle = 2.56$, which does not change with burst duration or fluence.
This observed anti-correlation would imply that magnetar bursts are not self-similar in burst duration: shorter bursts are not simply shorter copies of longer bursts. In the latter case, the power law index would be the same, but shifted to higher frequencies. Shorter bursts are also not simply shorter snapshots of the same process. Instead, it implies that the relevant variability time scales in each burst depend on the duration of the burst. Longer bursts are variable over a larger range of time scales and variable at higher frequencies. 
Exactly how this difference between short and long bursts manifests, and what implications it might have for magnetar burst emission models, is unclear. Again, more nuanced methods are required to better understand the broadband variability in magnetar bursts. Understanding this variability, in turn, is valuable for performing QPO searches with more precision than it is currently possible.
 
 \section{Theoretical discussion}
 
The QPOs in the tail of the giant flare from SGR 1900+14 lie in the range 28-155 Hz \citep{2005ApJ...632L.111S}. For the tail of the giant flare from SGR 1806-20, there are several QPOs in the range 18-150 Hz, and two isolated higher frequency signals at 625 Hz and 1840 Hz \citep{2005ApJ...628L..53I,2006ApJ...637L.117W,2006ApJ...653..593S}. Widths (FWHM) are in the range ~1 to ~20 Hz.

The most plausible explanation advanced for the giant flare QPOs is that they represent global seismic oscillations of the star, and it was swiftly realised that this would be a novel means of constraining not only the interior field strength (which is hard to measure directly) but also the dense matter equation of state \citep{2007MNRAS.374..256S,2007MNRAS.379L..63W}. The question of mode identification is therefore crucial. 

In the original discovery papers, the QPOs were tentatively identified with torsional shear modes of the neutron star crust and torsional Alfv\'en modes of the highly magnetized fluid core.   These identifications were based on the expected mode frequencies, which are set by both the size of the resonant volume and the relevant wave speed.  For crustal shear modes, the appropriate speed is the shear speed $v_s = (\mu_s/\rho)^{1/2}$ where $\mu_s$ is the shear modulus and $\rho$ the density. The shear modulus is of the order of the Coulomb potential energy $\sim Z^2e^2/r$ per unit volume $r^3$, where $r\sim(\rho/Am_p)^{-1/3}$ is the inter-ion spacing, while $Z$ and $A$ are the effective atomic number and mass number, respectively, of the ions in the crust. Using the shear modulus computed by \citet{Strohmayer91} and scaling by typical values for the inner crust \citep{Douchin01}, the shear velocity as shown by \citet{Piro2005} is:  

\begin{eqnarray}
v_s 	& = & 1.1 \times 10^8 \mathrm{cm/s}  \left(\frac{\rho}{10^{14} \mathrm{g/cm}^3}\right)^{1/6} \left(\frac{Z}{38}\right) \\ 
&& \times \left(\frac{302}{A}\right)^{2/3} \left(\frac{1-X_n}{0.25}\right)^{2/3} \nonumber \, ,
\end{eqnarray}

where $X_n$ is the fraction of neutrons. This yields a rough estimate for the frequency for the fundamental crustal shear mode of $\nu \sim v_s/2\pi R$ = 18 (10 km/$R$) Hz.  Full mode calculations find similar values, but with additional dependencies on the mass and radius of the star due to relativistic effects (see for example \citet{Samuelsson07}):  and it is this dependence that makes the modes potentially powerful diagnostics of the dense matter equation of state \citep{Lattimer07}.  Many of the lower QPO frequencies could be explained as angular harmonics with no radial nodes, whilst the two highest frequencies in the SGR 1806-20 giant flare were identified as radial overtones of these crustal modes. 

For torsional Alfv\'en modes of the core, the appropriate wave speed is the Alfv\'en speed $v_A = B/\sqrt{4\pi\rho}$ where $B$ is the magnetic field strength, giving

\begin{equation}
v_A = 10^8 \mathrm{cm/s}  \left(\frac{B}{10^{16} \mathrm{G}}\right) \left(\frac{10^{15} \mathrm{g/cm}^3}{\rho}\right)^{1/2}
\end{equation}

This yields a very rough estimate for the frequency of the fundamental torsional Alfv\'en mode of $\nu \sim v_A/4R =$ 25 (10 km/$R$) Hz \citep{Thompson01}.  Note however that the value of the field strength $B$ in magnetar cores is highly uncertain, as is the appropriate value of the density $\rho$. In principle only the charged component ($\sim$ 5-10\% of the core mass) should participate in Alfv\'en oscillations, reducing $\rho$, however there are mechanisms associated with superfluidity and superconductivity that can couple the charged and neutral components, leading to additional mass-loading. As above, full mode calculations that take into account relativistic effects lead to additional dependencies on neutron star mass and radius (see for example \citet{Sotani2008}).  It should also be noted that the Alfv\'en modes constitute continua rather than a set of discrete frequencies, since the field lines within the core have a continuum of lengths. 
It has been suggested that the observed QPOs might be associated with a turning points in the continuum of Alfven modes \citep{Levin07,Sotani2008}.

In fact, for a star with a magnetar strength field, crustal vibrations and core vibrations should couple together on very short timescales \citep{Levin06,Levin07}.  Considering them in isolation, as described above, is therefore not appropriate.  The current viewpoint, based on more detailed modelling that takes into account the magnetic coupling between crust and core, is that the QPOs are in fact associated with global {\bf magneto-elastic axial (torsional) oscillations} of the star \citep{Glampedakis06,Andersson09,Steiner09,vanHoven11,vanHoven12, Colaiuda11,Colaiuda12, Gabler12, Gabler13, Passamonti13a, Passamonti13b, Glampedakis14}.

Since magneto-elastic oscillations depend on the same physics described above, albeit now in a coupled system, they have frequencies in the same broad range as the simple estimates given above.  The freqencies are set by many factors including the dense matter equation of state (which sets mass and radius), field strength and geometry, superfluidity, superconductivity, and crust composition. Current magneto-elastic torsional oscillation models have had some success in explaining the presence of oscillations at frequencies of 155 Hz and below. However they struggle to explain the presence of the highest frequency oscillations, which should damp very rapidly \citep{vanHoven12, Gabler12}. Various solutions to this problem are under investigation, including coupling to polar modes \citep{Lander10, Lander11, Colaiuda12}, and resonances between crust and core that might develop as a result of superfluid effects \citep{Gabler13, Passamonti13b}. However until these issues are resolved, precise identification of the giant flare frequencies with specific global magneto-elastic modes remains a challenge.
 
The detection of frequencies in the smaller flares provides an entirely new viewpoint on this very difficult theoretical problem. It is therefore important to compare the properties of the giant flare QPOs to those detected in this study. We begin with the frequencies detected by averaging together multiple bursts from highly active episodes, at 93 Hz and 127 Hz. These frequencies are in the range found in the giant flares (indeed the strongest frequency found in the SGR 1806-20 giant flare was at 92 Hz). The widths are also comparable to the range observed in the tails of the giant flares. It therefore seems plausible that they are instances of the same phenomenon. If these frequencies do indeed represent global magneto-elastic oscillations the implication is that such vibrations are excited not only by giant flares, but also by trains of shorter bursts. This is important information for future theoretical studies of mode excitation.
 
The 260 Hz signal identified in one of the individual bursts is rather different. It is found in a frequency range where no signals were found in the giant flares. It is much broader than any of the oscillations seen in the tails of the giant flares, and has very high fractional amplitude. Whether it is the same phenomenon as was observed in the giant flares is therefore far from clear. If it is the same phenomenon, and we are seeing a magneto-elastic oscillation mode, then a detection in this frequency range would be valuable. Magneto-elastic oscillation models, as outlined above, have difficulty in explaining the lifetime of higher frequency signals in the giant flares.  This detection, with a much lower coherence, in a burst whose duration is comparable to the predicted lifetimes, provides a fresh perspective on this problem.
 
We may however be seeing something quite different. The giant flares consist of a short impulsive spike, followed by a long decaying tail as a trapped pair-plasma fireball slowly evaporates. In the smaller bursts, it is not clear whether a fireball forms: what we see may be more analogous to the impulsive spike of the giant flares. The variability that we have found in the short bursts in SGR 1550 (particularly the 260 Hz signal that appears to differ in properties) may instead be associated directly with the burst trigger process, be that a magnetospheric instability, or the yielding of the crust. It is worth noting that there are tentative claims of variability at 43 Hz in the first 200ms of the 1979 giant flare from SGR 0526-66 (Barat et al. 1983) and at 50 Hz in the first 500ms of the SGR 1806-20 giant flare (Terasawa et al. 2006, Geotail paper). However timing analysis of the peaks of the giant flares is heavily affected by dead time and saturation. In this respect the smaller flares, which are typically not saturated, may be more useful despite the lower countrates. However the variability that would be expected in the initial trigger and yielding phase of a magnetar burst has not been studied in detail. There are nonetheless plausible mechanisms that might lead to quasi-periodic behaviour. 

If the burst trigger is magnetospheric,  there may arise via plasma instabilities associated with magnetic reconnection (see for example Kliem, Karlický, \& Benz 2000).  If instead the trigger is crustal yielding, local effects and resonances may be significant.  It is not clear, for example, whether locally excited shear waves would immediately couple to the entire crust (and from there to the core) rather than resonating, with low Q-value, in a smaller cavity that is temporarily coupled very poorly to the rest of the crust during the yielding process.  Such local resonances would have quite different frequencies to those of global magneto-elastic oscillations.   More detailed theoretical studies of the trigger process will be required to resolve both this question, and the length of time required to establish global modes of any kind.  However the possibility that that the 260 Hz signal is a new and direct signature of the trigger process is an exciting one.

\section{Conclusions}

We have searched 263 bursts from SGR J1550-5418 for QPOs. We find one candidate QPO in the individual bursts searches. The signal is broad, but highly significant, and not close to any frequency observed in the giant flares. It is unclear whether this signal could come from the same phenomenon as the QPOs observed in the giant flares, or whether it may be associated directly with the burst trigger process.

Searching averaged periodograms reveals a significant signal at $\sim 10$ Hz in an averaged periodogram of all bursts with durations $\mathrm{T90} > 0.5$. This signal is comparable to the characteristic duration of a magnetar burst $\mathrm{T90}_\mathrm{max} = 0.1 \, \mathrm{s}$, but may be due to an inadequate broadband model for the periodogram. We find evidence for QPOs in periodograms averaging bursts from individual triggers, which are unlikely due to effects of averaging together bursts with vastly different timing properties. Two of these signals are located at $93\, \mathrm{Hz}$, where QPOs in the giant flares have been observed. The third is at a higher frequency, $127 \, \mathrm{Hz}$. We consider these signals to be the best candidates of neutron star oscillations from short magnetar bursts to date. All three signals can be interpreted in the framework of magneto-elastic oscillations invoked to explain QPOs in magnetar giant flares. The possibility that not only giant flares, but smaller flares may excite these oscillations also provides an important new piece of information for future theoretical studies of mode excitation.
For averaged periodograms, we can put constraints on weak signals that could have been there and would likely have been missed by our methods. For all but the lowest frequencies, our sensitivity to QPOs is lower than the observed fractional rms amplitudes in the giant flares. This is especially prominent for the high-frequency QPOs observed at $625$ Hz and higher. We thus conclude that except for the signals at $93$ Hz and $127$ Hz, there are no giant-flare like QPOs in this sample of small bursts.

Here we also characterised overall burst variability for the first time. We find a correlation between power-law index and burst duration. This implies that longer SGR bursts are variable over a broader range of time scales than short bursts, and are not simply longer versions of the short bursts. 
Further work is required to disentangle overall variability in magnetar bursts. This is unlikely to be possible with Fourier methods, but would be very rewarding both in terms of understanding emission mechanisms as well as untangling possible QPO signals from the overall burst morphology.

\acknowledgments
The authors thank Phil Uttley and Yuri Levin for useful discussions, and the referee for helpful suggestions.
D.H., CdA and ALW acknowledge support from a Netherlands Organization for Scientific Research (NWO) Vidi Fellowship (PI A. Watts).  
C.K. was partially supported by NASA grant NNH07ZDA001-GLAST. This publication is part of the GBM/Magnetar Key Project (NASA grant NNH07ZDA001-GLAST; PI: C. Kouveliotou).
AJvdH acknowledges support from the European Research Council via Advanced Investigator Grant no. 247295 (PI: R.A.M.J. Wijers).

\bibliography{sgr1550_references}
\bibliographystyle{apj}

\end{document}